# Temperature dependent Resonant X-ray Inelastic Scattering at Ni $L_3$-edge for $NaNiO_2$ and $LiNiO_2$.

Quentin Jacquet[a*], Kurt Kummer[b], Marie Guignard[f], Elisa Grépin[c,d,e], Sathiya Mariyappan[c,d,e], Nicholas B. Brookes[b], Sandrine Lyonnard[a]

[a] *Univ. Grenoble, Alpes, CEA, CNRS, Grenoble INP, IRIG, SyMMES, F-38000 Grenoble, France*

[b] *European Synchrotron Radiation Facility, F-38043 Grenoble Cedex, France*

[c] *Chimie du Solide-Energie, UMR 8260, Collège de France, 75231 Paris Cedex 05, France*

[d] *Sorbonne Université, 4 Place Jussieu, 75005, Paris, France*

[e] *Réseau sur le Stockage Electrochimique de l'Energie (RS2E), FR CNRS 3459, France*

[f] *Univ. Bordeaux, CNRS, Bordeaux INP, ICMCB, UMR 5026, Pessac, F-33600, France*

*Corresponding: quentin.jacquet@cea.fr*

**Abstract:**

$LiNiO_2$ is a promising cathode material for Li-ion battery but its atomic and electronic structure is under debate. Indeed, two sets of Ni-O distances are identified from local structural probes that are related with either Jahn-Teller distortion or bond disproportionation of $NiO_6$ octahedra. Moreover, $LiNiO_2$ undergoes a monoclinic to rhombohedral transition at 200 K which origin is still unclear. On the other hand, isostructural $NaNiO_2$ shows differences from LiNiO2, as it is a well-known Jahn-Teller distorted system, and it undergoes monoclinic to rhombohedral transition at 500 K associated to the loss of the Jahn-Teller distortion. To understand better these differences, we report here Ni $L_3$-edge Resonant inelastic X-ray scattering experiments on $LiNiO_2$ and $NaNiO_2$ at different temperatures (25 to 520 K) and follow the spectral changes below and above the phase transition temperatures. The observed RIXS spectra and the mapping indicate strong spectral changes for $NaNiO_2$ confirming the disappearance of Jahn-Teller distortion during phase transition while the changes are minor for LiNiO2 suggesting very few modifications in the local structure. Theoretical simulations of RIXS spectra are required for further understanding, however, we believe that the reported dataset can be a crucial resource for developing advanced simulations that are essential to deepening our understanding of the atomic and electronic structure of these nickelates.

**Introduction:**

LiNiO$_2$ (LNO) is currently receiving a regained interest in the field of Li-ion batteries as cathode material due to its high capacity[1]. However, LiNiO$_2$ atomic and electronic structure is still highly debated in both battery and solid state physics community[2–4]. Indeed, local atomic probes such as X-ray absorption spectroscopy (XAS) and neutron pair distribution function (PDF) analysis showed that at room temperature, LiNiO$_2$ features two sets of distinct Ni-O bond distances while long range diffraction techniques find regular symmetric NiO$_6$ octahedra with only one Ni-O distance[5,6]. This can be explained by the absence of long range ordering of a local NiO$_6$ distortion[6,7]. The debate regards how the long and short Ni-O bonds are distributed at the local scale and hence the resulting electronic structure. It is extremely challenging to measure experimentally non-periodic spatial distribution of bond distances. Therefore LiNiO$_2$ structure has been mostly investigated theoretically. Two main theories have been proposed: non-cooperative Jahn-Teller distortion of NiO$_6$ octahedra with a formal oxidation state +III (3d$^7$), or non-cooperative bond and charge disproportionation of Ni (+III) into Ni (+II) and Ni (+IV)[3,6]. Recent theoretical work using molecular dynamic simulations have also explained the non-cooperative distortions by either dynamic reorientations of the Jahn-Teller distortion or dynamic charge disproportionation[2,8,9]. To unpick the different theories, new experimental work including X-ray spectroscopies have been reported but more experimental work is still needed to reach a clear consensus [5,10].

One approach to experimentally measure a complex, potentially dynamic structure is to monitor structural parameters as function of temperature. Indeed, structural ordering might appear at low temperature helping to understand higher temperature phases. LiNiO$_2$ stabilises with monoclinic phase at low temperatures and undergoes a monoclinic to rhombohedral transition when heating above approx. 200 K showing some influence of the temperature on the atomic structure. However, the transition is very subtle since it only corresponds to β angle change from 90.11° to 90° [9]. Moreover, local distortion of the NiO$_6$ octahedra is still present over the phase transition as demonstrated by Chung et al. using variable temperature neutron PDF[6]. Several theories have been proposed to explain the increase of symmetry with temperature such as a disordered to ordered phase transition[2,6] or displacive phase transition (loss of the local distortion in the high temperature phase)[9]. The nature of the phase transition is still not completely clear, hence the need for further experimental investigations.

Second approach is to play with the composition, and compare the structural characteristics of LiNiO$_2$ to isoelectronic NaNiO$_2$ (NNO) or AgNiO$_2$ (ANO). NaNiO$_2$ and AgNiO$_2$ feature at room temperature a cooperative Jahn-Teller distortion and a cooperative charge ordering, respectively, both determined using classical diffraction methods[11–13]. NaNiO$_2$ also undergoes a monoclinic to rhombohedral phase transition when heating above approx. 460 K[6,14]. Nagle-Cocco et al. used diffraction, PDF and XAS at different temperatures across the transition and showed that the Jahn-Teller distortion disappears at the local scale [15], and hence that this transition is displacive. Comparing the temperature behaviour of NaNiO$_2$ and LiNiO$_2$ can provide interesting information to understand the phase transition in LiNiO$_2$.

A third approach is to use a large variety of techniques. Along that line, Ni L$_3$-edge RIXS has been recently used to understand the structure of Ni rare earth perovskites because it allows to directly measure d-d transitions which are very sensitive to the local atomic and electronic structure. Using this method combined with double cluster calculations, Lu et al. were able to quantify bond disproportionation of Ni sites in NdNiO$_3$[16]. Bisogni et al. measured the variable temperature d-d transition energy of NdNiO$_3$ and showed that this compound is a negative charge transfer material, hence that its Ni local electronic structure is better described by d$^8$L compared to d$^7$ (L being an electron hole on the ligands)[17]. Our team has recently reported Ni L$_3$-edge RIXS of LiNiO$_2$ and NaNiO$_2$[5].

We found that some d-d transitions in LNO were very comparable to NiO suggesting the presence of symmetric environments and a negative charge transfer character for Ni-O bond. NaNiO$_2$ RIXS spectra are different from LNO, with notably broader peaks in the d-d transition region indicative of the Jahn-Teller distortion in this material.

To probe further the atomic and electronic structure of LiNiO$_2$, we studied here the variable temperature Ni L$_3$-edge RIXS spectra of LiNiO$_2$ and NaNiO$_2$ above and below the monoclinic to rhombohedral transition. The novelty of this work is the report of the RIXS spectral evolution during heating and RIXS maps (measuring RIXS spectra for different excitation energies) at two different temperatures for both compounds. Strong changes in NaNiO$_2$ RIXS maps are observed during the disappearance of the Jahn-Teller distortion at high temperature showing that RIXS is indeed sensitive to the local structural changes of interest here. Conversely, LiNiO$_2$ RIXS spectra do not show large changes across the phase transition suggesting only weak modification of the local structure. Without advanced simulations, it is hard to extract precise physical parameters from these spectra. However, we believe that the reported dataset can serve as key input to develop advanced simulations needed to unlock our understanding of the complex atomic and electronic structure of these nickelates.

**Methods:**

Material synthesis aspects: LiNiO$_2$ is phase pure, commercial (BASF) and possess only 2% of Ni in the Li layer. More details on this material can be found in previous characterisation reports[5,18,19]. Two different NaNiO$_2$ samples were used, sample1 for the room temperature measurements, and sample2 for the high temperature measurements. NaNiO$_2$ synthesis conditions is reported elsewhere[20]. Both samples have the same crystal structure and cell parameters namely a = 5.3213 and 5.3226 Å (Δa = 0.02%), b = 2.8525 Å and 2.8497 Å (Δb = 0.1%), c = 5.5823 Å and 5.582 Å (Δc = 0.05%), β = 110.39 and 110.43° (Δβ = 0.03%). Moreover, 5% and 7% of NiO is found in the sample1 and sample2 using Rietveld refinement, respectively.

Variable temperature in situ XRD: Two experiments were performed. First, NaNiO$_2$ pristine was heated up to 525 K and cooled down to room temperature. The second experiment consisted in re-heating the powder up to 975 K and cooling it down again to room temperature. Both experiments were performed in an Xpert diffractometer (Panalytical/Malvern) with an Xcelerator detector using a Co Kα1-Kα2 source (λ = 1.7889 Å) in the 2θ-range 10°-80° with a 0.016° step size. The NaNiO$_2$ powder was loaded into an alumina crucible in a glove-box. The alumina crucible was quickly placed in an Anton Paar HTK 1200 furnace filled with helium to prevent any reaction of the mixture with the atmosphere. The furnace was then evacuated with the secondary vacuum pump. The pressure inside the furnace varies between 2 and 4 x 10$^{-4}$ mbar during the whole experiment. The diffraction patterns were recorded every 25 K for 1 hour, using a heating rate of 5K/min and a waiting time of 15 minutes to ensure temperature homogenization within the sample before the acquisition of the patterns. The total time of the experiment was 13 hours. For the second heat treatment, the powder resulting from the previous experiment was left in the alumina crucible in the furnace under vacuum. The pressure inside the furnace varies between 5 x 10$^{-4}$ and 2 x 10$^{-4}$ mbar during the whole experiment. The diffraction patterns were recorded every 50 K starting from 375 K for 1 hour, using a heating rate of 5°C/min and a waiting time of 15 minutes to ensure temperature homogenization within the sample before the acquisition of the patterns. The total time of the experiment was 21 hours.

Resonant inelastic X-ray scattering: The high-energy resolution RIXS measurements were carried out at the ID32 beamline of the ESRF [21]. Low temperature measurements are performed directly on LiNiO$_2$ and NaNiO$_2$ electrode glued on a Cu sample holder using Ag paste. Electrodes were either commercial (LNO – BASF) or homemade (NaNiO$_2$). An electrode is a composite structure composed of

LiNiO$_2$/NaNiO$_2$ particles embedded into a polymeric matrix (Polyvinylidene fluoride - PVDF) filled with conductive carbon (Carbon Super P – Csp). This composite is coated on Al foil leading to a self-standing flexible film easy to manipulate. High temperature measurements are performed on NaNiO$_2$ and LiNiO$_2$ powders pressed into a pellet, and glued onto the sample holder with Ag paste. Note that sample storage and operation was performed in an Ar filled glovebox. After preparation, sample holders were transferred into ID32 instrument vacuum chamber using an air-tight container and pumped for the night before introduction in the measurement chamber. During the high temperature measurement, the sample plate is heated and its temperature measured using a thermocouple. The temperature was varied between 300 and 550 K with 30 min to reach thermal equilibrium. For each temperature, a XAS spectrum and a RIXS spectrum with 853.05 eV excitation energy is measured. XAS spectrum quality is severely influenced by the absence of charge compensation for these sample holders. To measure the XAS and RIXS, the incident photon energy was tuned to the L$_3$ absorption edge of Ni and scanned across the edge (~850 - 860 eV) while resolving in energy in emitted photons. The energy resolution of the emitted photon was set to ~25 meV with a beam spot of 40 μm×3μm. The scattering angle was 149.5° and 30° grazing emission angle. No evolution of RIXS spectra was observed as a function of measurement time showing the absence of beam damage. Data analysis was performed using the RIXStoolbox[22].

**Results:**

**High Temperature XRD of NaNiO$_2$ under secondary vacuum**

RIXS experiments are conducted under vacuum conditions at up to 525 K, which might lead to LiNiO$_2$ and NaNiO$_2$ reduction. LiNiO$_2$ is stable under argon (which is a reductive atmosphere) up to 875 K,[23], it is hence very likely stable at the mild temperature of the RIXS experiment. However, less thermal stability data is available for NaNiO$_2$. Therefore, to check the stability of NaNiO$_2$ at high temperature under vacuum, we performed *in situ* XRD (Figure 1a). NaNiO$_2$ was placed into an alumina crucible under secondary vacuum (2 to 4 x 10$^{-4}$ mbar), and the diffraction patterns are recorded while heating up and after cooling down. The onset of the monoclinic to rhombohedral transition is observed at 425 K and full conversion is achieved at 450 K. Rhombohedral phase is the main phase up to 625 K. From 625 K to 875 K the 20° peak (003 reflection) intensity fades, and an unknown phase is present along with with NiO. From 875 K and up to 975 K, NiO is the main phase. Therefore, NaNiO$_2$ reduction under vacuum only appears at temperature higher than 600 K. This is further confirmed looking at the XRD pattern after cooling (Figure 1b). Room temperature (RT) XRD pattern of NaNiO$_2$ after heating up to 525 K shows a mixture of monoclinic and rhombohedral phase without NiO diffraction signal while the RT pattern of NaNiO$_2$ heated up to 975 K is mostly composed of NiO peaks. Note that the co-existence of monoclinic and rhombohedral phases during cooling could be due to the cooling process. Rietveld refinements of the variable temperature XRD patterns is conducted to extract the cell parameter evolution with temperature (Figure 1c). The interlayer distance increases linearly with temperature without major effect of the monoclinic → rhombohedral phase transition. At the contrary, the phase transition impacts the intralayer distance (Ni-Ni distance). The variable temperature XRD evolution of NaNiO$_2$ is generally consistent with previous reports from Nagle-Coco et al[15]. However, two minor differences are observed. Indeed, the authors reported a phase transition temperature of 460 K (compared to 425 K here) and they observed an abrupt change of interlayer distance across the phase transition (which is not the case here). The differences could originate from the atmosphere (vacuum versus argon) or the presence of NiO (5% impurity in our work) however further work should be performed to understand the origin of these slight discrepancies.

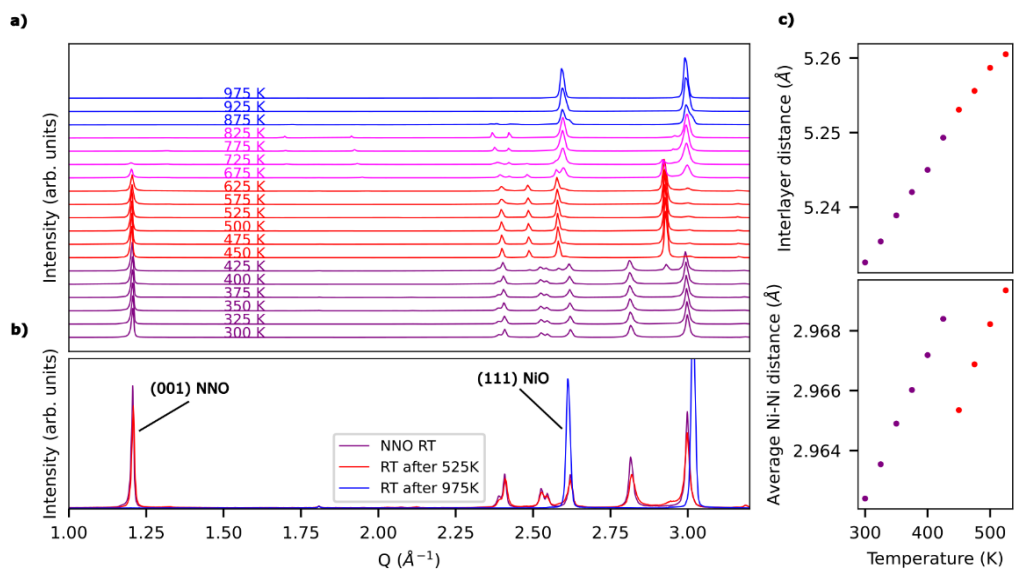

Figure 1: a) Variable temperature *in situ* XRD of NaNiO$_2$ under vacuum conditions. b) XRD patterns of pristine NaNiO$_2$, and after 525 K and 975 K heat treatment protocol, in purple, red and blue, respectively. c) Interlayer distance and average Ni-Ni distance in NaNiO$_2$ determined from the cell parameters as a function of temperature. Interlayer distance is determined using $c_{rhombohedral}/3$ and $C_{monoclinic} \times \sin(\beta_{monoclinic})$ for the rhombohedral and monoclinic cells, respectively, in red and purple.

**Variable temperature RIXS**

Ni L$_3$-edge total fluorescence yield (TFY) XAS spectra of LiNiO$_2$ and NaNiO$_2$ pellets on the high-temperature sample holder were measured and are shown in Figure 2a. We observe the traditional Ni L$_3$-edge shape for oxides having Ni in the formal +III oxidation state, which is composed of two peaks of relatively equal intensity at 852 and 854 eV (called hereafter peak A and B). [4,5,24] Before heating up the oxides to 520 K using temperature steps of 50 K, RIXS spectra were measured for both samples with an excitation energy of 853.05 eV. A spectrum in similar conditions was measured after the heat treatment in order to evaluate the possible reduction of the oxide at high temperature under ultra-high vacuum. LiNiO$_2$ RIXS spectra before and after heat treatment are identical, while a very small difference is observed in the case of NaNiO$_2$ (Figure 2b). TFY XAS spectra are also very similar before and after heat treatment with only a small increase of Peak A intensity suggesting a small partial surface reduction of the oxides (Figure S1). This is in agreement with the work of Dong et al. who reported NaNiO$_2$ reduction at 570 K[25] in air and our *in situ* XRD study.

Having confirmed the overall temperature stability of both samples, RIXS spectra ($E_{excit}$ = 853.05 eV) during heat treatment are measured (Figure 2c). For LiNiO$_2$, the room temperature spectrum shows an intense asymmetric peak at -1.2 eV and a broader peak at – 2 eV. The first peak is asymmetric because it is composed of two peaks clearly observed on the 25 K spectra. Upon heating, the spectra qualitatively remains similar with only a general broadening of the peaks. For NaNiO$_2$, at room temperature, the spectra is also composed of two peaks at -1.2 eV and -2 eV. Up to 450 K, the spectra remains similar and the peaks broaden. Between 450-500 K, the peak profile changes substantially as three peaks are now clearly visible at -1, -1.5 eV and -3 eV. This is the temperature of the monoclinic to rhombohedral transition hence showing that the loss of local Jahn-Teller distortion has a clear signature on the RIXS spectra. Now, we move to a more quantitative description of the spectra with temperature by fitting the spectra with Gaussian peaks (Figure S2, S3 and Table S1, S2). For NaNiO$_2$, six and five Gaussians are needed to fit the low and high temperature data, respectively (Figure 2d,e). Gaussian peak positions barely change in the 300-450 K temperature range. At 500 K, the five Gaussian positions are 0, -1, -1.5, -2.6, -3.6 eV which is significantly different compared to lower temperature. For LiNiO$_2$, six Gaussians are needed at 0, -1.26, -1.48, -1.7, -3, -4 eV (for 25 K) and their positions barely changes from 25 K to 520 K (Figure 2f). Indeed, the spectral change is mostly due to a broadening of all the peaks as the temperature increases with a very small shift of the -1.26 and -1.48 eV peaks to -1.20 and -1.45 eV respectively. Overall, we can conclude that the loss of Jahn-Teller transition is clearly observable on the RIXS spectra of NaNiO$_2$. In light of the weak changes observed in the RIXS spectra for LiNiO$_2$, this data suggest that there is very little modification of Ni environment in LiNiO$_2$ with temperature.

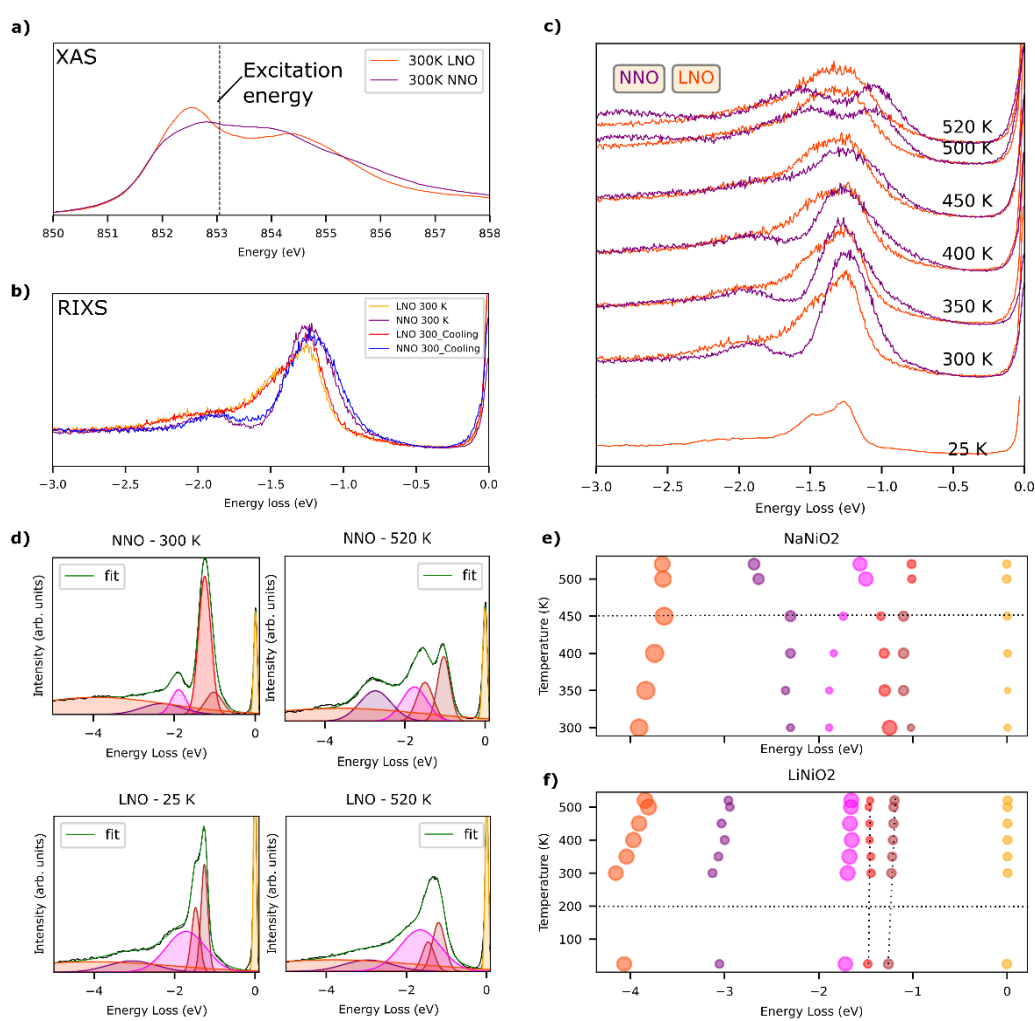

**Figure 2: (a)** TFY XAS Ni L$_3$-edge of LiNiO$_2$ and NaNiO$_2$ measured at 300 K using the high temperature sample holder. Dash line is showing the excitation energy used for the RIXS measurements shown in this figure. **(b)** RIXS spectra for LiNiO$_2$ and NaNiO$_2$ measured at 300 K before and after the variable temperature measurement. Excitation energy is 853 eV. **(c)** Ni L$_3$-edge RIXS spectra of LiNiO$_2$ and NaNiO$_2$ while ramping up the temperature. Note that RIXS spectra were acquired 30 min after the sample reached the desired temperature. **(d)** Gaussian fits of the RIXS spectra collected from NNO at 300 and 520 K and for LNO at 25 K and 520 K. Black curve is the data, and green is the fit almost

overlapping with the data. Six Gaussians are used and each peak is represented in a different colour from yellow to purple. (ef) Gaussian peak positions as a function of temperature for LNO and NNO. Size of the marker is proportional to the peak area. Vertical dashed lines are guide to the eye showing the evolution of the low energy loss peak. Horizontal dashed lines indicate phase transitions.

**RIXS maps at room and high temperature**

To gain more insights into the atomic and electronic structure change with temperature, energy loss spectra with different excitation energies were measured at 25 K and 520 K for $LiNiO_2$ and 300 K and 520 K for $NaNiO_2$ (Figure 3). First, we focus on the low temperature map for $LiNiO_2$. To quantify the spectral changes, each energy loss spectra has been fitted using six Gaussians (Figure 3a and Table S3-6). Low energy loss peaks at 0, -1.18 eV, -1.21 eV and -1.86 eV (yellow, brick, red and magenta colors) have an almost constant energy loss as a function of the excitation energy. The peak at 0 eV corresponds to the elastic scattering. The other peaks are called Raman-like because their energy loss is invariant with the excitation energy. They usually correspond to local transitions[17]. The nature of the low energy Raman-like signal in NiO is well established[26]. The first transition at -1.1 eV, is $^3T_{2g}$, and its energy corresponds to the crystal field splitting ($e_g^6 t_{2g}^2 \rightarrow e_g^5 t_{2g}^3$). The second transition at –1.8 eV involves crystal field together with spin flipping transition. By analogy with NiO, we hypothesize that the nature of the Raman-like signals observed at approx. -1.2 eV and -1.8 eV are d-d transitions. In $LiNiO_2$, there are at least two transitions around -1.2 eV, which is consistent with the presence of two different Ni-O bonds observed by local atomic structural probes. Discriminating between Jahn-Teller distortion and bond disproportionation is difficult at this stage without theoretical calculations. Higher energy loss spectra is usually composed of Raman-like charge transfer (CT) transitions (electrons for the oxygen ligand is involved in the emission process) and fluorescence-like transitions involving delocalized electrons. Energy loss of fluorescence-like transitions (FY) is proportional to the excitation energy as highlighted by dotted line in Figure 3a, while the charge transfer is mostly localized above -4 eV with a higher intensity at high excitation energy (orange Gaussian Figure 3a). High temperature RIXS map of $LiNiO_2$ is very similar to the low temperature one, with notably the presence of two d-d transition peaks around -1.1 eV. This suggests that two Ni-O environments are still present at this temperature. The shift of the d-d transitions from -1.2 eV to -1.1 eV can be explained by the reduction of the crystal field splitting due to the thermal expansion of the Ni-O distances. Moving on to 300 K $NaNiO_2$ RIXS map, the d-d transitions, CT and fluorescence transitions are still observed with some differences compared to $LiNiO_2$. First, the energy of the d-d transitions are different. $NaNiO_2$ feature peaks at -1.0 eV, -1.3 eV and -1.92 eV showing $NaNiO_2$ and $LiNiO_2$ have a different local structure. Second, the FY signal is less intense. At 520 K, the RIXS map changes substantially with the presence of only two low energy d-d transition peaks at -1 eV and -1.6 eV, and the appearance of a Raman-like peak at -2.72 eV. Moreover, the CT transition at low excitation energy (<853.5 eV) has almost no intensity. Finally, a new Raman-like peak appears (854.25 eV, - 3 eV). Qualitatively, we may attribute the absence of a doublet peak at – 1 eV to the loss of local Jahn-Teller distortion. Overall, the RIXS maps confirm that RIXS is sensitive to the loss local of Jahn-Teller distortion in $NaNiO_2$ and show that for $LiNiO_2$ Ni environment undergoes very little change across the monoclinic to rhombohedral transition suggesting that the two materials have a different behaviour.

**Discussion. The use of Ni $L_3$-edge RIXS in the battery community.** We have observed strong changes in the Ni $L_3$-edge RIXS spectra of $NaNiO_2$ across the monoclinic to rhombehdral transition consistent with the loss Jahn-Teller distortion. For $LiNiO_2$, the variation is weak suggesting that these two materials have different local structures. Beyond atomic structure, Ni $L_3$-edge RIXS has been key to understand the electronic structure of rare earth nickelates, and in particular the nature of the Ni-O bonds with the presence of holes on the oxygen atoms[17]. The charge compensation mechanism between Ni and O in battery materials is a highly debated topic because the use of oxygen as a redox center changes the paradigm of cathode material design[27–29]. In the battery community, O K-edge RIXS has been measured to understand to role of oxygen. Molecular $O_2$ was observed and proposed as a fingerprint for the participation of oxygen in the redox process by some authors[30]. More recent work show that the $O_2$ observed could actually originate from beam interaction with the sample[31] hence calling for new robust measurements. Ni $L_3$-edge RIXS is never measured in the battery community therefore we believe that the dataset reported in this work will be key in advancing our understanding of these complex oxide materials. However, interpreting RIXS requires adequate simulations. There are RIXS simulation software available such as Quanty[32] which are parameterized based on Density Functional Theory (DFT) calculations. However, $LiNiO_2$ is a very complex material to simulate due to the strong electron correlation and the disordered nature of its local structure, and hence there is wide range of different theory level predicting different properties for $LiNiO_2$[10,33–36]. Simulations groups need to find a consensus of the adequate models to describe $LiNiO_2$ atomic and electronic structure and we believe that this dataset can be key to prove or refute scientific hypothesis on the material or on the simulation methods themselves.

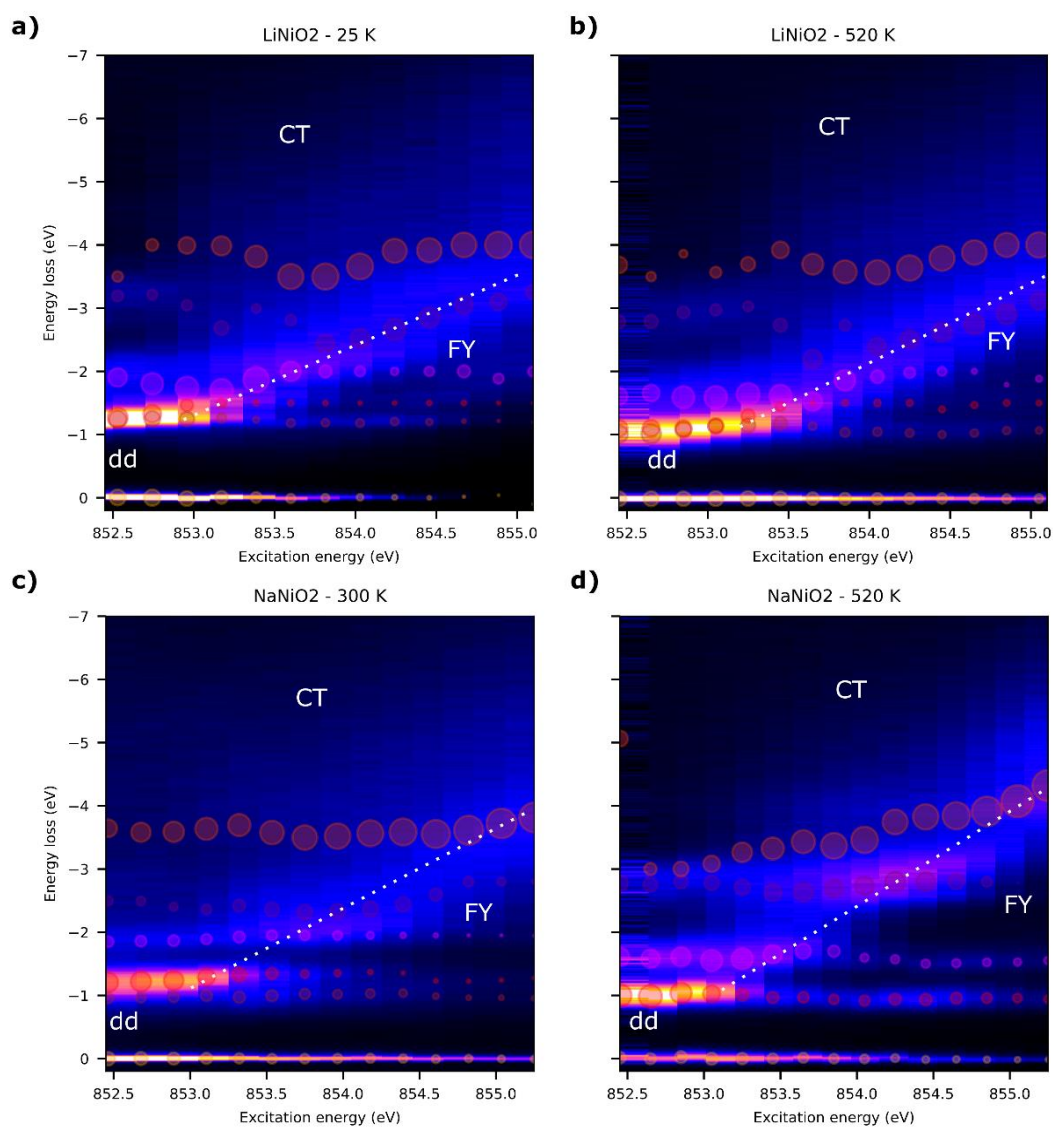

**Figure 3:** RIXS maps for LiNiO$_2$ at 25 K (a), LiNiO$_2$ 520 K (b), NaNiO$_2$ 300 K (c), NaNiO$_2$ at 520 K(d) showing the energy loss vs excitation energy. On each map, a white dotted line corresponding to the fluorescence-like signal (FY) is added. The regions for d-d and charge transfer (CT) transitions are also indicated. Moreover, the maps are superimposed with the result of the Gaussian fits represented as circles. The position of the circle corresponds to the peak position, while its size is proportional to the Gaussian area.

**Conclusion:**

In this work, we have reported variable temperature XRD of NaNiO$_2$ and variable temperature Ni L$_3$-edge RIXS of NaNiO$_2$ and LiNiO$_2$. Temperatures are chosen below and above the monoclinic to rhombohedral transitions reported for both compounds. The XRD data confirms the presence of the monoclinic to rhombohedral transition for NaNiO$_2$ together with the absence of reduction under vacuum up to 625 K. Variable RIXS spectra for NaNiO$_2$ changes substantially between 300 K and 520 K. Two peaks are observed in the crystal field splitting energy domain (-1.2 eV) at low temperature consistent with the presence of Jahn-Teller distortion. At 520 K, this doublet becomes a single peak consistent with the reported loss of local Jahn-Teller distortion. For LiNiO$_2$, a doublet is also observed at 25 K in agreement with the presence of two Ni-O bonds distances in this material. However, at 520 K the doublet is still present showing that in contrast to NaNiO$_2$, the monoclinic to rhombohedral transition occurs with minor change in the Ni local environment in LiNiO$_2$. The RIXS maps contain several dispersive and non-dispersive features that change with temperature, indicating a complex electronic structure change. Key information on the local and electronic structure, and in particular the role of the Ni and O in the charge compensation mechanism, could be extracted from these spectra using simulations. This was demonstrated by the physics community when studying the rare earth nickelates. However, the level of theory needed to simulate LiNiO$_2$ is still under debate. Therefore, we believe that this dataset opens possibilities for further work and constitutes a valuable experimental contribution to help the development of RIXS simulations needed to deepen the understanding of oxides.

**Acknowledgement:** Beamtime at the ESRF was granted within the Battery Pilot Hub MA4929 "Multi-scale Multi-techniques investigations of Li-ion batteries: towards a European Battery Hub". The data is accessible using the following DOI : doi.org/10.15151/ESRF-DC-2008703222. We acknowledge support from EU H2020 project BIG-MAP (grant agreement ID: 957189).

**Author contribution:** Q.J. conceived the idea. S.L. supervised the project. M.G., E.C., S.M. prepared the samples, M.G. performed the XRD measurements. K.K., N.B. aligned and set the beamline. Q.J. K.K., N.B. performed the beamtime and analysed the results. Q.J wrote the manuscript, all authors revised the manuscript. All authors discussed the results and contributed to the manuscript.

Supporting information

**Temperature dependent Resonant X-ray Inelastic Scattering at Ni $L_3$-edge for NaNiO$_2$ and LiNiO$_2$.**

Quentin Jacquet[a*], Kurt Kummer[b], Marie Guignard[f], Elisa Grépin[c,d,e], Sathiya Mariyappan[c,d,e], Nicholas B. Brookes[b], Sandrine Lyonnard[a]

[a] *Univ. Grenoble, Alpes, CEA, CNRS, Grenoble INP, IRIG, SyMMES, F-38000 Grenoble, France*

[b] *European Synchrotron Radiation Facility, F-38043 Grenoble Cedex, France*

[c] *Chimie du Solide-Energie, UMR 8260, Collège de France, 75231 Paris Cedex 05, France*

[d] *Sorbonne Université, 4 Place Jussieu, 75005, Paris, France*

[e] *Réseau sur le Stockage Electrochimique de l'Energie (RS2E), FR CNRS 3459, France*

[f] *Univ. Bordeaux, CNRS, Bordeaux INP, ICMCB, UMR 5026, Pessac, F-33600, France*

*Corresponding: quentin.jacquet@cea.fr*

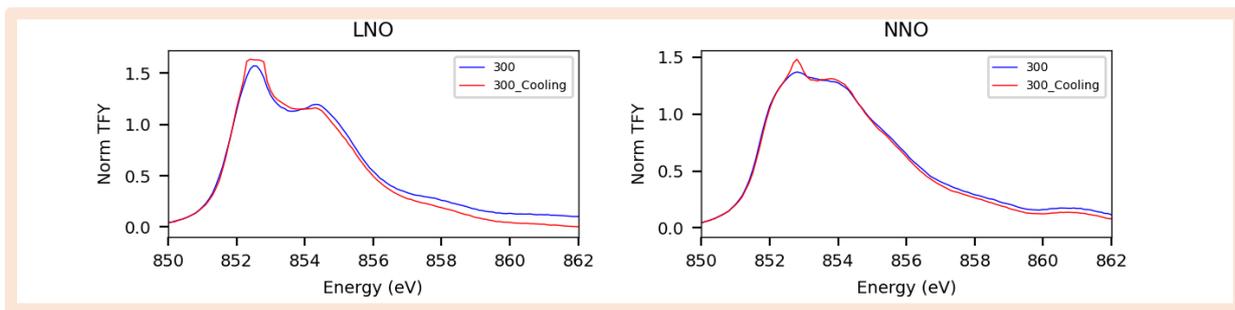

Figure S1: Total Fluoresence yield X-ray absorption spectra of LNO and NNO before and after the heat treatment. Both materials feature very slight shift of the spectra towards lower energy. In particular, Peak A at 852.5 eV is slightly more intense after heating after heating. This is probably due a small reduction of the oxide at the surface while heating. Note that the shape of the XAS spectra is distorted due to charging effects.

|  | I1 | p1 | s1 | I2 | p2 | s2 | I3 | p3 | s3 | I4 | p4 | s4 | I5 | p5 | s5 |
|---|---|---|---|---|---|---|---|---|---|---|---|---|---|---|---|
| **300 K** | 0.005 | 0.004 | 0.040 | 0.007 | -1.230 | 0.168 | 0.001 | -1.895 | 0.162 | 0.000 | -2.334 | 0.345 | 0.001 | -3.640 | 1.896 |
| **350 K** | 0.003 | 0.005 | 0.041 | 0.005 | -1.257 | 0.185 | 0.001 | -1.918 | 0.149 | 0.001 | -2.369 | 0.302 | 0.001 | -3.544 | 1.866 |
| **400 K** | 0.006 | 0.004 | 0.040 | 0.004 | -1.246 | 0.225 | 0.001 | -1.903 | 0.159 | 0.001 | -2.367 | 0.375 | 0.001 | -3.557 | 1.881 |
| **450 K** | 0.006 | 0.001 | 0.044 | 0.003 | -1.241 | 0.253 | 0.001 | -1.841 | 0.214 | 0.001 | -2.398 | 0.500 | 0.001 | -3.617 | 1.864 |
| **500 K** | 0.006 | -0.005 | 0.052 | 0.002 | -1.012 | 0.150 | 0.003 | -1.500 | 0.334 | 0.001 | -2.638 | 0.486 | 0.001 | -3.653 | 2.000 |
| **520 K** | 0.005 | -0.004 | 0.057 | 0.002 | -1.015 | 0.144 | 0.003 | -1.562 | 0.315 | 0.001 | -2.687 | 0.445 | 0.001 | -3.662 | 2.000 |

Table S1: Gaussian fit parameters for variable temperature NaNiO$_2$

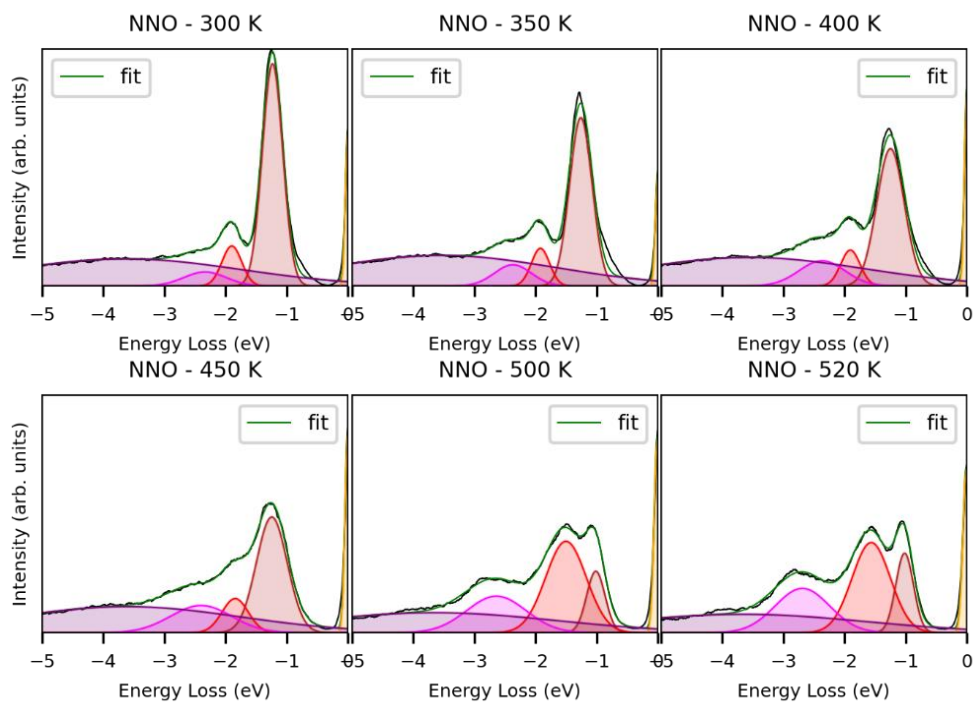

Figure S2: Fit of the variable temperature RIXS spectra for NaNiO$_2$

| | I1 | p1 | s1 | I2 | p2 | s2 | I3* | p3 | s3 | I4 | p4 | s4 | I5 | p5 | s5 | I6 | p6 | s6 |
|---|---|---|---|---|---|---|---|---|---|---|---|---|---|---|---|---|---|---|
| 25 K | 0.010 | -0.003 | 0.037 | 0.005 | -1.260 | 0.085 | 0.003 | -1.479 | 0.101 | 0.002 | -1.718 | 0.491 | 0.001 | -3.055 | 0.600 | 0.000 | -4.065 | 2.008 |
| 300 K | 0.008 | 0.004 | 0.038 | 0.004 | -1.229 | 0.100 | 0.002 | -1.445 | 0.126 | 0.002 | -1.694 | 0.511 | 0.001 | -3.128 | 0.600 | 0.000 | -4.154 | 2.063 |
| 350 K | 0.008 | 0.004 | 0.038 | 0.003 | -1.221 | 0.110 | 0.002 | -1.446 | 0.129 | 0.002 | -1.674 | 0.508 | 0.001 | -3.065 | 0.600 | 0.000 | -4.040 | 2.056 |
| 400 K | 0.008 | 0.004 | 0.039 | 0.003 | -1.213 | 0.124 | 0.002 | -1.461 | 0.129 | 0.002 | -1.648 | 0.499 | 0.001 | -2.998 | 0.600 | 0.000 | -3.968 | 2.066 |
| 450 K | 0.008 | 0.005 | 0.039 | 0.002 | -1.207 | 0.146 | 0.001 | -1.462 | 0.144 | 0.002 | -1.668 | 0.517 | 0.001 | -3.032 | 0.600 | 0.001 | -3.907 | 2.068 |
| 500 K | 0.008 | 0.004 | 0.040 | 0.002 | -1.206 | 0.151 | 0.001 | -1.470 | 0.140 | 0.002 | -1.658 | 0.507 | 0.000 | -2.945 | 0.600 | 0.001 | -3.806 | 2.033 |
| 520 K | 0.008 | 0.004 | 0.041 | 0.002 | -1.195 | 0.160 | 0.001 | -1.454 | 0.149 | 0.002 | -1.655 | 0.504 | 0.001 | -2.959 | 0.600 | 0.001 | -3.843 | 2.069 |

Table S2: Gaussian fit parameters for variable temperature LiNiO$_2$. i3 is not fitted but set as i3 = i2*0.60.

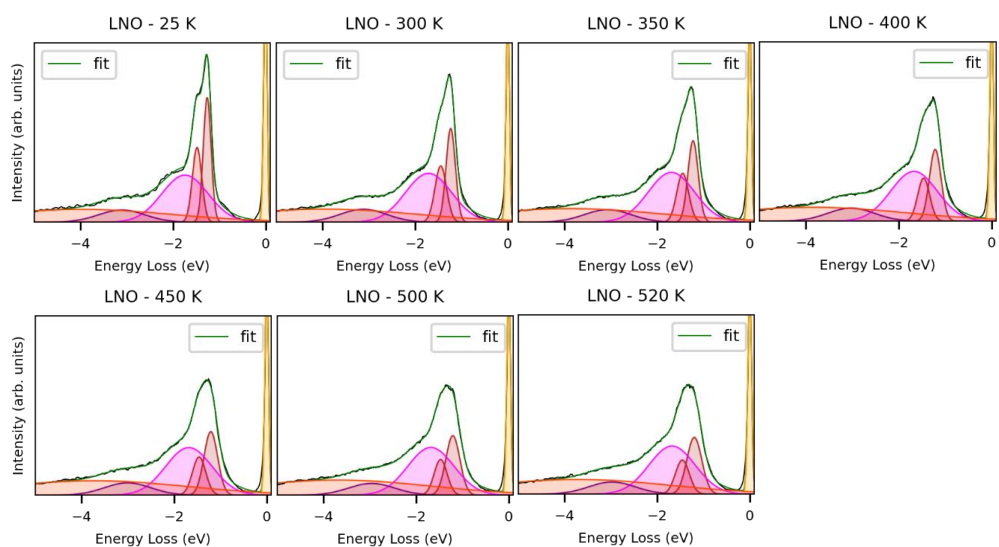

Figure S3: Fit of the variable temperature RIXS spectra for LiNiO$_2$

| Energy | I1 | p1 | s1 | I2 | p2 | s2 | I3 | p3 | s3 | I4 | p4 | s4 | I5 | p5 | s5 | I6 | p6 | s6 |
|---|---|---|---|---|---|---|---|---|---|---|---|---|---|---|---|---|---|---|
| 852.45 | 0.0077 | 0.0160 | 0.0500 | 0.0043 | -1.0088 | 0.1341 | 0.0026 | -1.1000 | 0.2000 | 0.0017 | -1.6043 | 0.2989 | 0.0007 | -2.7827 | 0.4456 | 0.0002 | -3.6845 | 2.2617 |
| 852.64 | 0.0077 | 0.0146 | 0.0500 | 0.0041 | -1.0133 | 0.1581 | 0.0025 | -1.1000 | 0.2000 | 0.0017 | -1.6582 | 0.2902 | 0.0007 | -2.7884 | 0.4642 | 0.0001 | -3.5000 | 1.5150 |
| 852.83 | 0.0073 | 0.0152 | 0.0500 | 0.0038 | -1.0767 | 0.1176 | 0.0023 | -1.1000 | 0.2000 | 0.0019 | -1.5899 | 0.4500 | 0.0008 | -2.9336 | 0.3312 | 0.0002 | -3.8611 | 0.7496 |
| 853.01 | 0.0077 | 0.0132 | 0.0500 | 0.0032 | -1.1299 | 0.1203 | 0.0019 | -1.1408 | 0.2000 | 0.0019 | -1.5893 | 0.4605 | 0.0007 | -2.9731 | 0.3397 | 0.0002 | -3.5673 | 1.0855 |
| 853.20 | 0.0076 | 0.0116 | 0.0500 | 0.0026 | -1.1448 | 0.1425 | 0.0016 | -1.2972 | 0.2000 | 0.0018 | -1.6368 | 0.5108 | 0.0005 | -3.0278 | 0.3732 | 0.0003 | -3.6976 | 1.1229 |
| 853.39 | 0.0066 | 0.0111 | 0.0500 | 0.0018 | -1.1787 | 0.1663 | 0.0011 | -1.4452 | 0.1405 | 0.0018 | -1.6134 | 0.4762 | 0.0005 | -2.7308 | 0.6000 | 0.0004 | -3.9211 | 1.1743 |
| 853.58 | 0.0055 | 0.0131 | 0.0500 | 0.0006 | -1.1329 | 0.2000 | 0.0004 | -1.4965 | 0.0995 | 0.0016 | -1.5207 | 0.3928 | 0.0010 | -2.1987 | 0.6000 | 0.0006 | -3.6957 | 1.2575 |
| 853.77 | 0.0049 | 0.0159 | 0.0500 | 0.0008 | -1.0348 | 0.2000 | 0.0005 | -1.5000 | 0.1585 | 0.0018 | -1.8601 | 0.4014 | 0.0006 | -2.7271 | 0.3494 | 0.0007 | -3.5744 | 1.2910 |
| 853.96 | 0.0046 | 0.0150 | 0.0500 | 0.0007 | -1.0432 | 0.1926 | 0.0004 | -1.5000 | 0.2000 | 0.0010 | -1.9157 | 0.3031 | 0.0010 | -2.3963 | 0.5039 | 0.0008 | -3.5657 | 1.3078 |
| 854.14 | 0.0040 | 0.0124 | 0.0500 | 0.0007 | -1.0253 | 0.1929 | 0.0004 | -1.5000 | 0.1942 | 0.0008 | -1.9792 | 0.2577 | 0.0012 | -2.5107 | 0.4822 | 0.0009 | -3.6354 | 1.3212 |
| 854.33 | 0.0038 | 0.0112 | 0.0500 | 0.0006 | -1.0010 | 0.1757 | 0.0004 | -1.3985 | 0.2000 | 0.0005 | -2.0000 | 0.2676 | 0.0012 | -2.6238 | 0.5205 | 0.0009 | -3.7878 | 1.2795 |
| 854.52 | 0.0034 | 0.0135 | 0.0500 | 0.0006 | -1.0227 | 0.1782 | 0.0004 | -1.4659 | 0.2000 | 0.0002 | -2.0000 | 0.2588 | 0.0013 | -2.7439 | 0.5328 | 0.0008 | -3.8807 | 1.2666 |
| 854.71 | 0.0030 | 0.0171 | 0.0500 | 0.0006 | -1.0629 | 0.1859 | 0.0003 | -1.5000 | 0.1924 | 0.0001 | -1.7890 | 0.2217 | 0.0013 | -2.8993 | 0.5525 | 0.0008 | -4.0000 | 1.2731 |
| 854.90 | 0.0027 | 0.0131 | 0.0500 | 0.0004 | -1.0634 | 0.2000 | 0.0003 | -1.5000 | 0.2000 | 0.0002 | -1.8833 | 0.3468 | 0.0011 | -3.1123 | 0.4966 | 0.0009 | -4.0000 | 1.3118 |
| 855.09 | 0.0025 | 0.0107 | 0.0500 | 0.0003 | -1.0608 | 0.1995 | 0.0002 | -1.5000 | 0.1481 | 0.0002 | -1.8121 | 0.2413 | 0.0008 | -3.2875 | 0.4778 | 0.0009 | -4.0000 | 1.3423 |
| 855.27 | 0.0019 | 0.0090 | 0.0500 | 0.0003 | -1.0510 | 0.2000 | 0.0002 | -1.5000 | 0.1339 | 0.0002 | -1.8496 | 0.1994 | 0.0006 | -3.5311 | 0.3982 | 0.0011 | -4.0000 | 1.2582 |
| 855.65 | 0.0015 | 0.0039 | 0.0500 | 0.0002 | -1.0483 | 0.1898 | 0.0001 | -1.5000 | 0.1999 | 0.0001 | -1.8184 | 0.1788 | 0.0008 | -3.9649 | 0.4807 | 0.0008 | -4.0000 | 1.4374 |

Table S3: Fitting parameters for the 520 K RIXS map of LiNiO$_2$

| Energy | l1 | p1 | s1 | l2 | p2 | s2 | l3 | p3 | s3 | l4 | p4 | s4 | l5 | p5 | s5 | l6 | p6 | s6 |
|---|---|---|---|---|---|---|---|---|---|---|---|---|---|---|---|---|---|---|
| 852.10 | 0.0070 | 0.0030 | 0.0500 | 0.0078 | -1.1860 | 0.0725 | 0.0047 | -1.2128 | 0.1514 | 0.0018 | -1.8686 | 0.3138 | 0.0008 | -3.1154 | 0.3462 | 0.0001 | -3.5000 | 1.3663 |
| 852.30 | 0.0074 | -0.0133 | 0.0500 | 0.0081 | -1.2160 | 0.0702 | 0.0049 | -1.2421 | 0.1363 | 0.0018 | -1.8573 | 0.3451 | 0.0008 | -3.1401 | 0.3296 | 0.0001 | -3.5000 | 1.4094 |
| 852.50 | 0.0087 | -0.0060 | 0.0500 | 0.0077 | -1.2405 | 0.0697 | 0.0046 | -1.2711 | 0.1512 | 0.0017 | -1.9040 | 0.3501 | 0.0008 | -3.1934 | 0.2849 | 0.0002 | -3.5000 | 1.1377 |
| 852.70 | 0.0092 | -0.0021 | 0.0500 | 0.0061 | -1.2490 | 0.0712 | 0.0037 | -1.3316 | 0.1470 | 0.0016 | -1.8026 | 0.5034 | 0.0006 | -3.2146 | 0.2897 | 0.0003 | -4.0000 | 0.9389 |
| 852.90 | 0.0076 | 0.0124 | 0.0500 | 0.0044 | -1.2463 | 0.0853 | 0.0026 | -1.4659 | 0.1016 | 0.0018 | -1.7358 | 0.5152 | 0.0004 | -3.0470 | 0.4363 | 0.0004 | -3.9934 | 1.1557 |
| 853.10 | 0.0057 | -0.0079 | 0.0500 | 0.0022 | -1.2602 | 0.0740 | 0.0013 | -1.5000 | 0.1091 | 0.0019 | -1.7000 | 0.4663 | 0.0006 | -2.6853 | 0.5643 | 0.0006 | -3.9849 | 1.1858 |
| 853.30 | 0.0041 | -0.0059 | 0.0500 | 0.0009 | -1.2335 | 0.0767 | 0.0005 | -1.5000 | 0.1049 | 0.0021 | -1.8697 | 0.5218 | 0.0004 | -2.9948 | 0.3697 | 0.0007 | -3.8167 | 1.2337 |
| 853.50 | 0.0030 | 0.0106 | 0.0500 | 0.0010 | -1.1847 | 0.1314 | 0.0006 | -1.5000 | 0.1070 | 0.0019 | -2.0000 | 0.3629 | 0.0005 | -2.8116 | 0.4248 | 0.0008 | -3.5000 | 1.4423 |
| 853.70 | 0.0023 | 0.0013 | 0.0500 | 0.0008 | -1.1864 | 0.1356 | 0.0005 | -1.5000 | 0.0916 | 0.0011 | -2.0000 | 0.2544 | 0.0011 | -2.4385 | 0.5064 | 0.0009 | -3.5000 | 1.3969 |
| 853.90 | 0.0016 | -0.0067 | 0.0500 | 0.0008 | -1.1828 | 0.1493 | 0.0005 | -1.5000 | 0.0500 | 0.0008 | -2.0000 | 0.2462 | 0.0012 | -2.5256 | 0.5040 | 0.0009 | -3.6601 | 1.3341 |
| 854.10 | 0.0010 | -0.0063 | 0.0500 | 0.0009 | -1.2058 | 0.1255 | 0.0005 | -1.5000 | 0.0530 | 0.0004 | -2.0000 | 0.2887 | 0.0013 | -2.6608 | 0.5998 | 0.0007 | -3.9068 | 1.3544 |
| 854.30 | 0.0006 | 0.0023 | 0.0500 | 0.0007 | -1.1989 | 0.1084 | 0.0004 | -1.5000 | 0.1044 | 0.0004 | -2.0000 | 0.4383 | 0.0012 | -2.8500 | 0.6000 | 0.0007 | -3.9152 | 1.4094 |
| 854.50 | 0.0002 | -0.0172 | 0.0500 | 0.0004 | -1.2115 | 0.0686 | 0.0003 | -1.5000 | 0.1776 | 0.0005 | -2.0000 | 0.5303 | 0.0011 | -3.0532 | 0.5252 | 0.0008 | -4.0000 | 1.4365 |
| 854.70 | 0.0001 | -0.0407 | 0.0500 | 0.0002 | -1.2159 | 0.0635 | 0.0001 | -1.5000 | 0.2000 | 0.0005 | -1.8860 | 0.3542 | 0.0010 | -3.0977 | 0.4989 | 0.0008 | -4.0000 | 1.4734 |
| 854.90 | 0.0001 | 0.1000 | 0.0500 | 0.0001 | -1.1934 | 0.0505 | 0.0001 | -1.5000 | 0.1579 | 0.0006 | -1.9956 | 0.3273 | 0.0008 | -3.2569 | 0.4430 | 0.0009 | -4.0000 | 1.3497 |

Table S4: Fitting parameters for the 25 K RIXS map of LiNiO$_2$

| Energy | I1 | p1 | s1 | I2 | p2 | s2 | I3 | p3 | s3 | I4 | p4 | s4 | I5 | p5 | s5 | I6 | p6 | s6 |
|---|---|---|---|---|---|---|---|---|---|---|---|---|---|---|---|---|---|---|
| 852.25 | 0.0097 | -0.0053 | 0.0369 | 0.0026 | -1.0109 | 0.1861 | 0.0025 | -1.2922 | 0.1467 | 0.0010 | -1.8604 | 0.2000 | 0.0004 | -2.5762 | 0.3177 | 0.0003 | -3.2768 | 1.7497 |
| 852.45 | 0.0086 | 0.0020 | 0.0369 | 0.0010 | -0.9553 | 0.2000 | 0.0039 | -1.2090 | 0.1748 | 0.0010 | -1.8495 | 0.2000 | 0.0004 | -2.4966 | 0.4000 | 0.0004 | -3.6455 | 1.5496 |
| 852.65 | 0.0078 | -0.0026 | 0.0381 | 0.0008 | -0.9618 | 0.2000 | 0.0045 | -1.2201 | 0.1591 | 0.0011 | -1.8590 | 0.2000 | 0.0004 | -2.4943 | 0.4000 | 0.0004 | -3.5818 | 1.5905 |
| 852.85 | 0.0067 | -0.0012 | 0.0393 | 0.0009 | -0.9725 | 0.2000 | 0.0046 | -1.2368 | 0.1454 | 0.0011 | -1.8661 | 0.2000 | 0.0004 | -2.4038 | 0.3350 | 0.0005 | -3.5895 | 1.5225 |
| 853.05 | 0.0056 | -0.0045 | 0.0389 | 0.0012 | -0.9981 | 0.1933 | 0.0037 | -1.2910 | 0.1367 | 0.0010 | -1.8900 | 0.1899 | 0.0006 | -2.3620 | 0.3105 | 0.0006 | -3.6360 | 1.4744 |
| 853.25 | 0.0048 | -0.0043 | 0.0384 | 0.0013 | -0.9716 | 0.1861 | 0.0023 | -1.3292 | 0.1461 | 0.0012 | -1.9247 | 0.2000 | 0.0008 | -2.4237 | 0.3399 | 0.0006 | -3.6957 | 1.3571 |
| 853.45 | 0.0045 | 0.0016 | 0.0385 | 0.0012 | -0.9952 | 0.1834 | 0.0015 | -1.3485 | 0.1514 | 0.0011 | -1.9500 | 0.1794 | 0.0009 | -2.3624 | 0.3695 | 0.0007 | -3.5776 | 1.3690 |
| 853.65 | 0.0043 | 0.0011 | 0.0385 | 0.0010 | -1.0228 | 0.1916 | 0.0010 | -1.3382 | 0.1299 | 0.0010 | -1.9500 | 0.1740 | 0.0010 | -2.3175 | 0.3880 | 0.0008 | -3.4978 | 1.3678 |
| 853.85 | 0.0040 | -0.0016 | 0.0392 | 0.0008 | -1.0190 | 0.1903 | 0.0007 | -1.3537 | 0.1216 | 0.0009 | -1.9500 | 0.1614 | 0.0010 | -2.3361 | 0.3715 | 0.0008 | -3.5209 | 1.3512 |
| 854.05 | 0.0037 | -0.0054 | 0.0406 | 0.0007 | -1.0306 | 0.1997 | 0.0004 | -1.3695 | 0.1054 | 0.0006 | -1.9500 | 0.1601 | 0.0011 | -2.3688 | 0.3901 | 0.0009 | -3.5654 | 1.3000 |
| 854.25 | 0.0035 | -0.0022 | 0.0387 | 0.0006 | -1.0370 | 0.1963 | 0.0002 | -1.3539 | 0.1005 | 0.0003 | -1.9500 | 0.1753 | 0.0009 | -2.4427 | 0.4000 | 0.0010 | -3.5861 | 1.3092 |
| 854.45 | 0.0030 | 0.0059 | 0.0387 | 0.0004 | -0.9902 | 0.1736 | 0.0002 | -1.2303 | 0.1209 | 0.0002 | -1.9500 | 0.1546 | 0.0007 | -2.5943 | 0.4000 | 0.0011 | -3.5535 | 1.3072 |
| 854.65 | 0.0028 | -0.0023 | 0.0362 | 0.0003 | -0.9632 | 0.1648 | 0.0002 | -1.2252 | 0.1094 | 0.0001 | -1.9500 | 0.1144 | 0.0004 | -2.8000 | 0.4000 | 0.0012 | -3.6137 | 1.2631 |
| 854.85 | 0.0026 | 0.0039 | 0.0354 | 0.0003 | -0.9628 | 0.1741 | 0.0002 | -1.2420 | 0.1021 | 0.0001 | -1.9500 | 0.1000 | 0.0003 | -2.8000 | 0.4000 | 0.0013 | -3.7217 | 1.1986 |
| 855.05 | 0.0025 | 0.0060 | 0.0349 | 0.0002 | -0.9562 | 0.1879 | 0.0002 | -1.2754 | 0.1068 | 0.0001 | -1.9427 | 0.1000 | 0.0001 | -2.8000 | 0.4000 | 0.0014 | -3.8161 | 1.1214 |

Table S5: Fitting parameters for the 300 K RIXS map of NaNiO$_2$

| Energy | I1 | p1 | s1 | I2 | p2 | s2 | I3 | p3 | s3 | I4 | p4 | s4 | I5 | p5 | s5 |
|---|---|---|---|---|---|---|---|---|---|---|---|---|---|---|---|
| 852.45 | 0.0038 | -0.0154 | 0.0610 | 0.0072 | -1.0004 | 0.1314 | 0.0025 | -1.6083 | 0.2484 | 0.0015 | -2.7779 | 0.3051 | 0.0002 | -5.0641 | 2.0000 |
| 852.65 | 0.0038 | 0.0004 | 0.0561 | 0.0071 | -0.9853 | 0.1307 | 0.0025 | -1.5802 | 0.2383 | 0.0013 | -2.7558 | 0.2745 | 0.0002 | -3.0000 | 1.5782 |
| 852.85 | 0.0043 | -0.0236 | 0.0590 | 0.0060 | -1.0379 | 0.1306 | 0.0023 | -1.6095 | 0.2814 | 0.0012 | -2.8000 | 0.2746 | 0.0003 | -3.0001 | 1.4895 |
| 853.05 | 0.0046 | 0.0010 | 0.0578 | 0.0038 | -1.0268 | 0.1334 | 0.0023 | -1.5529 | 0.3432 | 0.0012 | -2.7802 | 0.3243 | 0.0003 | -3.0820 | 1.3739 |
| 853.25 | 0.0042 | -0.0014 | 0.0563 | 0.0019 | -1.0072 | 0.1378 | 0.0023 | -1.5796 | 0.3336 | 0.0012 | -2.7097 | 0.3749 | 0.0005 | -3.2637 | 1.2877 |
| 853.45 | 0.0034 | -0.0024 | 0.0548 | 0.0016 | -0.9670 | 0.1430 | 0.0022 | -1.6503 | 0.2673 | 0.0013 | -2.6263 | 0.3928 | 0.0007 | -3.3253 | 1.2098 |
| 853.65 | 0.0034 | -0.0139 | 0.0535 | 0.0015 | -0.9700 | 0.1403 | 0.0016 | -1.7000 | 0.2515 | 0.0014 | -2.6341 | 0.4220 | 0.0008 | -3.4333 | 1.1597 |
| 853.85 | 0.0031 | 0.0002 | 0.0523 | 0.0012 | -0.9381 | 0.1371 | 0.0009 | -1.7000 | 0.2770 | 0.0016 | -2.6933 | 0.3430 | 0.0010 | -3.3668 | 1.1810 |
| 854.05 | 0.0026 | 0.0036 | 0.0545 | 0.0014 | -0.9186 | 0.1329 | 0.0006 | -1.5954 | 0.2146 | 0.0020 | -2.7351 | 0.3024 | 0.0011 | -3.4605 | 1.1582 |
| 854.24 | 0.0021 | -0.0085 | 0.0557 | 0.0018 | -0.9400 | 0.1338 | 0.0006 | -1.5719 | 0.2142 | 0.0023 | -2.8000 | 0.2979 | 0.0011 | -3.7446 | 0.9832 |
| 854.44 | 0.0017 | 0.0136 | 0.0568 | 0.0018 | -0.9354 | 0.1352 | 0.0007 | -1.5033 | 0.1897 | 0.0021 | -2.8000 | 0.2978 | 0.0012 | -3.8263 | 0.9180 |
| 854.64 | 0.0015 | 0.0100 | 0.0530 | 0.0016 | -0.9569 | 0.1426 | 0.0008 | -1.5104 | 0.1611 | 0.0014 | -2.8000 | 0.3114 | 0.0014 | -3.8434 | 0.9165 |
| 854.84 | 0.0014 | 0.0162 | 0.0502 | 0.0012 | -0.9598 | 0.1475 | 0.0008 | -1.5233 | 0.1444 | 0.0006 | -2.8000 | 0.3332 | 0.0016 | -3.9110 | 0.9644 |
| 855.04 | 0.0011 | 0.0164 | 0.0477 | 0.0009 | -0.9586 | 0.1517 | 0.0006 | -1.5292 | 0.1563 | 0.0001 | -2.7816 | 0.0100 | 0.0018 | -4.0737 | 1.0157 |
| 855.24 | 0.0009 | 0.0152 | 0.0449 | 0.0008 | -0.9386 | 0.1442 | 0.0006 | -1.5522 | 0.1970 | 0.0001 | -2.6125 | 0.0100 | 0.0018 | -4.3199 | 0.9296 |
| 855.44 | 0.0007 | 0.0164 | 0.0452 | 0.0007 | -0.9240 | 0.1450 | 0.0005 | -1.5630 | 0.2001 | 0.0001 | -2.6000 | 0.0100 | 0.0019 | -4.5295 | 0.8678 |
| 855.64 | 0.0004 | 0.0194 | 0.0483 | 0.0007 | -0.9224 | 0.1408 | 0.0005 | -1.5762 | 0.1990 | 0.0001 | -2.6000 | 0.0818 | 0.0020 | -4.7413 | 0.8324 |

Table S6: Fitting parameters for the 520 K RIXS map of NaNiO$_2$